\documentclass[aps,prc,superscriptaddress,showpacs,nofootinbib,floatfix,twocolumn]{revtex4}

\usepackage{graphicx}
\usepackage{amssymb,amsmath}

\begin{document}

\title{Parity violating elastic electron scattering from $^{27}$Al and the Qweak measurement}
\author{C. J. Horowitz}\email{horowit@indiana.edu}
\affiliation{Department of Physics and Nuclear Theory Center, Indiana University, Bloomington, IN 47405, USA}
\date{\today}
\begin{abstract}
I calculate the parity violating asymmetry $A_{pv}$ for elastic electron scattering from $^{27}$Al in order to compare with the Qweak experiment ``background'' $^{27}$Al measurement.  I find that Coulomb distortions, and the quadrupole form factor, reduce $A_{pv}$ near the diffraction minimum.  The Qweak data can be used to confirm the neutron radius of $^{27}$Al, if nuclear structure uncertainties are indeed small, as I suggest, and one can estimate inelastic and impurity contributions.   This could provide an important check of the measurement, analysis, and theory.
\end{abstract}
\smallskip
\maketitle

\section{Introduction}
Parity violating elastic electron scattering from a heavy nucleus is interesting for several reasons.  First, it is sensitive to neutron distributions, because the weak charge of a neutron is much larger than that of a proton \cite{dds,bigprex}.  The PREX experiment at Jefferson Laboratory has pioneered using parity violation to measure the neutron density of $^{208}$Pb \cite{ref:prexI,ref:prexFF}.   This result will be improved with a follow on PREX-II measurement \cite{ref:prexII}, and the approved CREX experiment will measure the neutron radius of $^{48}$Ca \cite{ref:CREX,CREXworkshop}.  Note that neutron densities can also be measured, in principle, with elastic neutrino scattering \cite{1207.0693,hep-ex/0511042}, a process that could be important in astrophysics \cite{PRD68.023005}.

Second, parity violation can be used to test the standard model, \cite{9Be,12CSouder}.  For example, a precision electron scattering experiment could measure the weak charge of $^{12}$C.  This is like an atomic parity measurement without many of the atomic structure uncertainties.   Electron scattering could also probe radiative corrections, such as Coulomb distortions \cite{couldist}, and other nuclear structure effects \cite{don2013}.

Recently, the Qweak experiment is measuring the parity violating asymmetry $A_{pv}$  for electrons scattering from $^{27}$Al \cite{qweak,qweakexpt}.   This experiment primarily measures $A_{pv}$ from hydrogen to determine the weak charge of the proton and test the standard model.  However, the Qweak hydrogen target has $^{27}$Al windows.  These windows are an important source of background because $A_{pv}$ for $^{27}$Al is much larger than that for hydrogen.\footnote{Because the weak charge of $^{27}$Al is much larger than the small weak charge of the proton}  
Therefore a separate precise measurement of $A_{pv}$ for $^{27}$Al is being undertaken, in order to subtract the contribution of window scattering from the main hydrogen measurement.  

In this paper I calculate $A_{pv}$ for elastic scattering from $^{27}$Al in order to compare to the Qweak measurement.  While the experimental Qweak background subtraction does not depend directly on theory, our calculations will provide a check of some of  the important assumptions involved in the subtraction.   For example, if the $^{27}$Al measurement disagrees with theory, one could worry that this might be due to a common systematic error that could also impact the main hydrogen measurement.  Alternatively if the $^{27}$Al measurement agrees with theory, this will help support the validity of the background subtraction procedure.  

Finally the precision $^{27}$Al measurement is interesting in its own right and can be used to confirm the neutron radius $R_n$ of $^{27}$Al.  In this paper I calculate the effects of Coulomb distortions, discuss the main nuclear structure sensitivities, and present results for the cross section and parity violating asymmetry.
In Section \ref{Formalism} I present my formalism for elastic electron scattering from a non spin zero nucleus.  Section \ref{Results} presents results for  $A_{pv}$, while possible inelastic contributions are discussed in Section \ref{Inelastic}, and I conclude in Section \ref{Conclusions}.
          
\section{Formalism}
\label{Formalism}

I begin calculating the cross section, first in Born approximation and then including Coulomb distortions.  I describe $^{27}$Al as a very simple pure $d_{5/2}$ proton hole in a relativistic mean field model using the FSUgold \cite{ref:fsu} interaction.   Table \ref{Tab:radii} presents proton $R_p$, neutron $R_n$, and charge $R_{ch}$ radii of this model.  The proton density has a spherically symmetric part $\rho_p^0(r)$ and a part that contains the $d_{5/2}$ hole $\rho_p^2({\bf r})$,
\begin{equation}
\rho_p({\bf r})=\rho_p^0(r)+\rho_p^2({\bf r}).
\end{equation}
Explicitly I write out
\begin{equation}
\begin{split}
\rho_p^0(r)=\frac{1}{4\pi r^2} \Bigl\{ 
 2({G_{s_{1/2}}}^2+{F_{s_{1/2}}}^2)+ 4({G_{p_{3/2}}}^2+{F_{p_{3/2}}}^2) \\
 + 2({G_{p_{1/2}}}^2+{F_{p_{1/2}}}^2)+ 5({G_{d_{5/2}}}^2+{F_{d_{5/2}}}^2) \Bigr\} 
\end{split}
\label{rhop0}
\end{equation} 
where $G_j$ is the Dirac upper component and $F_j$ is the lower component of the proton wave functions for the occupied states $j=1s_{1/2},1p_{3/2},1p_{1/2},1d_{5/2}$ \cite{cjhbds}.  Note that there are only 5 protons in the $d_{5/2}$ level.  The contribution of the hole is written,
\begin{equation}
\rho_p^2({\bf r}) =\frac{1}{r^2}({G_{d_{5/2}}}^2+{F_{d_{5/2}}}^2)(\frac{1}{4\pi}-|Y_{2M}({\bf \hat r})|^2),
\end{equation}
where $Y_{2M}$ is a spherical harmonic.  These densities are normalized,
$\int d^3r \rho_p^0(r)=Z=13$, and $\int d^3r \rho_p^2({\bf r})=0$.

\begin{table}[h]
\caption{\label{Tab:radii} Proton $R_p$, neutron $R_n$, and charge $R_{ch}$ radii of $^{27}$Al for FSUgold RMF model.} 
%\begin{ruledtabular}
\begin{tabular}{*{3}{c}}
$R_p$  (fm) &$R_n$ (fm) & $R_{ch}$ (fm) \\
\hline
2.904 & 2.913 & 3.013\\
\end{tabular}
%\end{ruledtabular}
\end{table}

I assume that the neutron density is spherically symmetric,
$
\rho_n({\bf r})=\rho_n^0(r),
$
where $\rho_n^0(r)$ has the same form as Eq. (\ref{rhop0}), with 5 replaced by 6 and the Dirac wave functions are slightly different for the neutron states.  The normalization is $\int d^3r \rho_n^0(r)=N=14$.

The proton form factor $F_p(q)$ is, 
\begin{equation}
F_p(q)=\frac{1}{Z}\int d^3r \bigl(\rho_p^0(r)+\rho_p^2({\bf r})\bigr) {\rm e}^{i{\bf q \cdot r}},
\end{equation}
where $q$ is the momentum transfer.  
\begin{equation}
\begin{split}
F_p(q)= &\frac{1}{Z}\int d^3r \bigl(\rho_p^0(r)+\rho_p^2({\bf r})\bigr)\\
 & \bigl\{j_0(qr)-\sqrt{20\pi}j_2(qr)Y_{20}+\sqrt{36\pi}j_4(qr)Y_{40}\bigr\}.
 \end{split}
\end{equation}
Here I have expanded the plane wave and only kept terms that make nonzero contributions.  Finally, $j_L$ are spherical Bessel functions.  There are contributions for $L=0, 2, 4$.  Squaring  $|F_p|^2$ and averaging over orbital angular momentum projection $M$ from -2 to 2 yields,
\begin{equation}
|F_p|^2={C_0}^2+{C_2}^2+{C_4}^2,
\end{equation}
 with
 \begin{equation}
C_0(q)= \frac{1}{Z}\int d^3r \rho_p^0(r) j_0(qr),
\label{eq.c0}
\end{equation}
\begin{equation}
C_2(q)=\frac{1}{Z}\sqrt{\frac{10}{7}}\int_0^\infty dr\, j_2(qr)\bigl({G_{d_{5/2}}}^2+{F_{d_{5/2}}}^2\bigr),
\label{eq.c2}
\end{equation}
and
\begin{equation}
C_4(q)=\frac{1}{Z}\sqrt{\frac{2}{7}}\int_0^\infty dr\, j_4(qr)\bigl({G_{d_{5/2}}}^2+{F_{d_{5/2}}}^2\bigr).
\label{eq.c4}
\end{equation}
The form factor is normalized $F_p(0)=C_0(0)=1$.  The neutron form factor only has an $L=0$ contribution,
$
%\begin{equation}
F_n(q)= \frac{1}{N}\int d^3r \rho_n^0(r) j_0(qr),
%\end{equation}
$
and is also normalized $F_n(0)=1$.

\begin{figure}
\resizebox{0.5\textwidth}{!}{%
\includegraphics{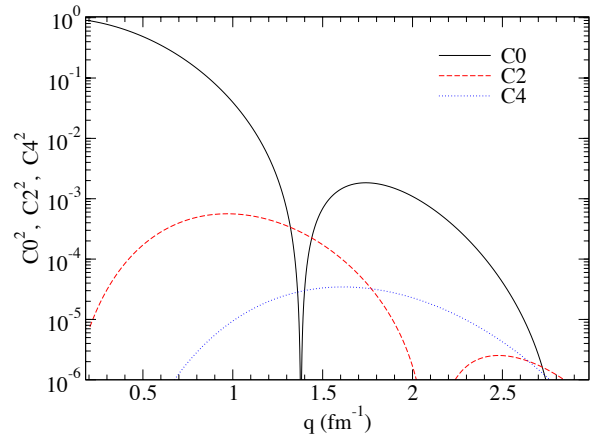}
}
\caption{(Color on line) Square of the proton form factors $C_0^2$ (solid), $C_2^2$ (dashed), and $C_4^2$ (dotted), Eqs. (\ref{eq.c0},\ref{eq.c2},\ref{eq.c4}), versus momentum transfer $q$.}
\label{Fig.1}
\end{figure}

In Figure \ref{Fig.1} I plot the square of the proton form factors for $L=0,2,4$.  I see that the $L=0$ contribution $C_0^2$ dominates, except near the diffraction minimum around $q=1.4$ fm$^{-1}$ where $C_2$ is important.  The $L=4$ contribution $C_4^2$ is small.  These results are similar to much earlier shell model calculations using harmonic oscillator wave functions \cite{old27Al}.

The cross section $d\sigma/d\Omega$ for elastic electron scattering can be calculated in Born approximation, see for example ref. \cite{old27Al}.
\begin{equation}
\frac{d\sigma}{d\Omega}=\sigma_m\eta\frac{q_\mu^2}{q^2}F_L^2 \approx \sigma_m F_L^2
\label{eq.sigma}
\end{equation}
Here the Mott cross section is,
$
%\begin{equation}
\sigma_m=\frac{Z^2\alpha^2\cos^2\theta/2}{4E_i^2\sin^4\theta/2},
%\label{eq.mott}
%\end{equation}
$
with $\theta$ the scattering angle and $E_i$ the incident electron energy.  The recoil factor is $\eta=(1+2E_i/M\sin^2\theta/2)^{-1}\approx 1$ with $M$ the mass of the nucleus.  Finally $q_\mu^2=q^2-(E_i-E_f)^2$ with $E_F$ the final electron energy and the three momentum transfer is $q^2=4E_iE_f\sin^2\theta/2+(E_i-E_f)^2$. 

The longitudinal form factor $F_L$ is $F_p$ folded with the electric form factor of a single proton $G_E$,
$F_L(q)^2=G_E(q)^2F_p^2$, however see ref. \cite{SO} for a discussion of spin-orbit currents.
Equation (\ref{eq.sigma}) neglects contributions from transverse currents.  For $^{27}$Al these have been calculated in ref. \cite{old27Al} and found to be very small, comparable to $C_4^2$ in Fig. \ref{Fig.1}.  Note that in general I do not expect large transverse contributions for the forward angle kinematics of Qweak.

Coulomb distortions can be important near diffraction minima.  If one neglects the aspherical $\rho_p^2({\bf r})$, Coulomb distortions have been calculated, in the usual way, by numerically solving the Dirac equation and summing partial waves \cite{couldist}.  Our procedure is to include Coulomb distortions for $C_0$ contributions exactly \cite{couldist} and then simply add $C_2$ contributions in Born approximation.  Our best estimate for the cross section is,
\begin{equation}
\frac{d\sigma}{d\Omega}\approx \frac{d\sigma}{d\Omega}(C_0)\Bigl|_{DW}+\sigma_m \xi^2C_2^2,
\label{eq.sigmafinal}
\end{equation}
where I have neglected $C_4$ and transverse contributions.   I add a parameter $\xi$ to include a generous estimate for uncertainties in the nuclear structure and for the effects of Coulomb distortions on the $C_2$ contribution.  Note that I do not expect Coulomb distortions to be very important for the $C_2$ contribution because it is only relevant for $q$ near 1.4 fm$^{-1}$, which is far away from the diffraction minimum in $C_2$.  Allowing $\xi^2$ to very between 0.5 and 1.5 provides a generous uncertainty range that likely includes many far more sophisticated nuclear structure models.   

\begin{figure}
\resizebox{0.5\textwidth}{!}{%
\includegraphics{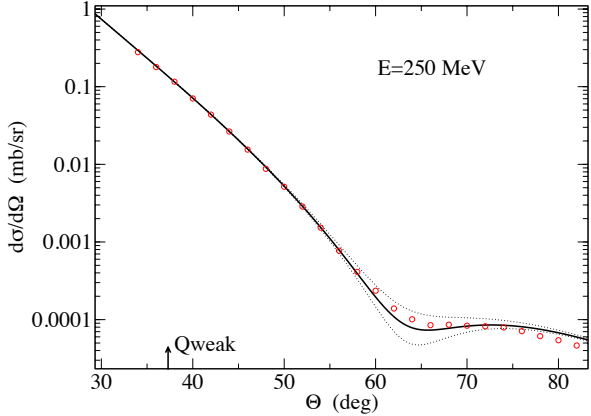}
}
\caption{(Color on line) Cross section for elastic scattering of 250 MeV electrons from $^{27}$Al versus laboratory scattering angle $\Theta$.  The red circles are experimental results from ref. \cite{27Alsigma} while the solid line is Eq. (\ref{eq.sigmafinal}) with $\xi^2=1$.  The upper (and lower) dotted lines correspond to $\xi^2=0.5$ (1.5) and indicates a very generous estimate for nuclear structure uncertainties.  The average momentum transfer of the Qweak experiment corresponds to the arrow near 37 degrees.}
\label{Fig.2}
\end{figure}

Equation (\ref{eq.sigmafinal}) is plotted in Fig. \ref{Fig.2} and agrees well with measured cross sections for 250 MeV elastic electron scattering from $^{27}$Al \cite{27Alsigma}.  This good agreement indicates that our FSUgold relativistic mean field model has approximately the correct charge radius and that our picture of $^{27}$Al as a simple $d_{5/2}$ proton hole is a reasonable first approximation.  Furthermore it suggests that values of $\xi^2 < 0.5$ or greater than 1.5 are unlikely because they disagree with measured cross sections.  Note that Eq. (\ref{eq.sigmafinal}) is slightly above the data for very large scattering angles beyond 80 degrees.  This suggests that our form factors may have slightly the wrong shape (probably surface thickness).   However these large momentum transfers are not relevant for the Qweak experiment.

 I now calculate the parity violating asymmetry $A_{pv}$.  This is the fractional cross section difference for scattering of electrons of positive helicity $+$ and negative helicity $-$.
 \begin{equation}
 A_{pv}=\frac{d\sigma/d\Omega|_+-d\sigma/d\Omega|_-}{d\sigma/d\Omega|_+ + d\sigma/d\Omega|_-}
 \end{equation}
Perhaps the simplest approximation is to assume $\rho_p^2=0$ and that the proton and neutron distributions have the same shape so that $F_p(q)=F_n(q)$.  In Born approximation $A_{pv}$ is then simply proportional to $q^2$,
\begin{equation}
A_{pv}=A_{pv}^0=-\frac{G_Fq^2 Q_W}{4\pi\sqrt{2}\alpha Z}.
\label{apv0}
\end{equation}
Here $G_F$ is the Fermi constant and the total weak charge of $^{27}$Al, $Q_W$ is,
\begin{equation}
Q_W=Q_nN+Q_pZ =-12.8919.
\end{equation}
The weak charge of the neutron $Q_n$ is -1 at tree level.  However including radiative corrections \cite{erleretal,radcorrec} I use $Q_n=-0.9878$.  The weak charge of the proton $Q_p$ is small.  It is $1-4\sin^2\theta_W$ at tree level and I use $Q_p=0.0721$ with radiative corrections.

Including different neutron and proton distributions, $A_{pv}$ in Born approximation becomes,
\begin{equation}
A_{pv}=A_{pv}^0\frac{C_0C_0^W+C_2C_2^W+C_4C_4^W}{{C_0}^2+{C_2}^2+{C_4}^2}.
\label{apvborn}
\end{equation}
Here the weak Coulomb form factors $C_L^W$ are Fourier transforms of the weak charge density $\rho_W({\bf r})$,
\begin{equation}
\rho_W({\bf r})\approx Q_n \rho_n(r)+Q_p\rho_p({\bf r}).
\end{equation}
If one  defines $C_0^W(q)=\int d^3r \rho_W({\bf r}) j_0(qr)/Q_W$ so that $C_0^W(0)=1$, I have,
\begin{equation}
C_0^W(q)=\frac{1}{Q_W}\Bigl[Q_nNF_n(q)+Q_pZC_0(q)\Bigr].
\end{equation}
The weak form factors for $L=2,4$ are small because the proton weak charge $Q_p$ is small and I assumed that the neutron density is spherically symmetric.  I have  $C_2^W(q)=Q_pZ C_2(q)/Q_W$ and $C_4^W=Q_pZC_4(q)/Q_W$.

It would be very interesting to calculate core polarization corrections to $C_2^W$.  The proton hole is expected to polarize the neutron density and this can make a significant contribution to $C_2^W$ because the weak charge of a neutron is much larger than that of a proton.  I think the net effect of this core polarization on $A_{pv}$ can be included by reducing the value of $\xi$, see below. 

I now include Coulomb distortions for the $C_0$ (and $C_0^W$) contributions as I did for the cross section.  I calculate $A_{pv}=A_{DW}(C_0)$ exactly for spherically symmetric weak charge and E+M charge distributions by solving the Dirac equation numerically for an electron moving in an axial vector weak potential (of order a few eV) and the Coulomb potential \cite{couldist}.  Then I add the $C_2$ and $C_2^W$ contributions in Born approximation.  My best estimate for $A_{pv}$ is,
\begin{equation}
A_{pv}\approx \frac{\frac{d\sigma}{d\Omega}(C_0)|_{DW}A_{DW}(C_0)+\sigma_m\xi^2C_2C_2^WA_{pv}^0}{\frac{d\sigma}{d\Omega}(C_0)|_{DW}+\sigma_m\xi^2{C_2}^2}.
\label{apvfinal}
\end{equation}
Note that the second term in the numerator is small because $C_2^W$ and $Q_p$ are small.  Therefore the primary impact of the $C_2$ contribution is to increase the denominator, and therefore reduce $A_{pv}$, for $q$ near the diffraction minimum in $C_0$.  Equation (\ref{apvfinal}) reproduces the exact distorted wave result if $C_2$ is small and reproduces the full Born approximation result, Eq. (\ref{apvborn}), when the effects of Coulomb distortions are small.

\section{Results}
\label{Results}

\begin{figure}
\resizebox{0.5\textwidth}{!}{%
\includegraphics{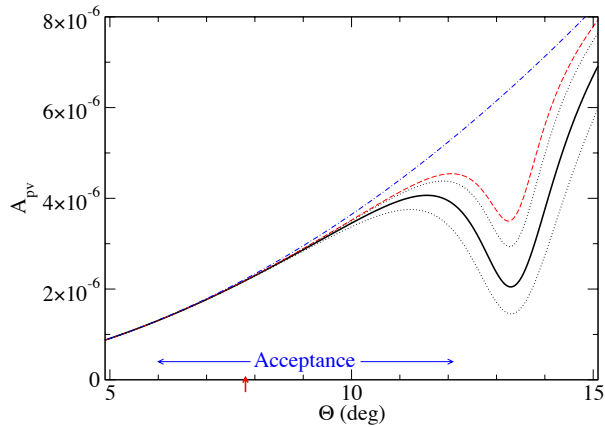}
}
\caption{(Color on line) Parity violating asymmetry $A_{pv}$ for elastic scattering of 1160 MeV electrons from $^{27}$Al versus scattering angle $\Theta$.  The blue dot dashed curve shows plane wave Born approximation results for the same shaped neutron and proton densities, $A_{pv}^0$ in Eq. (\ref{apv0}).  The red dashed curve shows full distorted wave results for spherically symmetric neutron and proton densities, $A_{DW}(C_0)$.  Distorted wave results for $C_0$ plus plane wave $C_2$ and $C_2^W$ contributions, Eq. (\ref{apvfinal}), are shown by the black upper dotted line, solid line, and lower dotted line  for $\xi^2=0.5$,1, and 1.5. The average momentum transfer of the Qweak experiment is shown by the red arrow, while the angular acceptance is very roughly indicated by the the blue arrows.}
\label{Fig.3}
\end{figure}

Figure \ref{Fig.3} shows $A_{pv}$ for electrons of energy 1160 MeV (the energy of Qweak) versus scattering angle.  Even for spherically symmetric neutron and proton densities, Coulomb distortions significantly reduce $A_{pv}$, in the diffraction minimum near 13 degrees, compared to the plane wave $A_{pv}^0$ result.  Including $C_2$ and $C_2^W$ contributions further reduces $A_{pv}$.  Our best estimate for $A_{pv}$ shown by the solid black line, Eq. (\ref{apvfinal}) is only one third of $A_{pv}^0$ in the minimum near 13.5 degrees.

The average momentum transfer of Qweak corresponds to about 7.8 degrees as shown by the red arrow in Fig. \ref{Fig.3}.  At this angle the effects of Coulomb distortions and $C_2$ are small.  This suggests that the final uncertainty in acceptance averaged theory results may be small.  However the angular acceptance of the Qweak experiment is large, as shown very roughly by the blue arrows in Fig. \ref{Fig.3}, and includes some acceptance near the diffraction minimum.  The cross section falls very rapidly with increasing angle so that only a small fraction of the events may come from angles near the diffraction minimum.  Therefore, the acceptance averaged contribution of the large dip in $A_{pv}$ may not be large.  Nevertheless it is important to carefully average our $A_{pv}$ predictions, with its complex shape, with the experimental acceptance.  Note that our $A_{pv}$ is not proportional to $q^2$.  Indeed for angles beyond 11.5 degrees, $A_{pv}$ actually decreases with increasing $q^2$.  Therefore one should be somewhat careful in extrapolating a measurement at one $q^2$ to a different $q^2$.

The asymmetry $A_{pv}$ is somewhat sensitive to nuclear structure uncertainties, for scattering angles beyond about 11 degrees.  This is shown in Fig. \ref{Fig.3} by the dotted error bands which correspond to different $\xi^2$values.  However, this nuclear structure uncertainty is very small at the average $q^2$ near 7.8 degrees.  Therefore the remaining nuclear structure uncertainty, by the time one averages over the acceptance, may be small.  This should be carefully checked.

\section{Inelastic Contributions}
\label{Inelastic}

My calculation of $A_{pv}$ can be compared to the Qweak measurement.  However, there are important inelastic backgrounds that need to be estimated before one can fully interpret the experimental results.  The Qweak spectrometer has only modest energy resolution and accepts inelastically scattered electrons with energy losses up to about 100 MeV.  Therefore one will also have contributions from discrete excited states, collective giant resonances, and quasielastic scattering.  For the forward angle Qweak kinematics I expect the discrete excited states to be dominated by Coulomb multipoles.  For these one can easily make an estimate of $A_{pv}$, see also ref. \cite{bigprex}.  The most important property is the isospin of the excitation.  Isoscalar excitations, where neutrons move in phase with protons, should have $A_{pv}\approx A_{pv}^0$, see Eq. (\ref{apv0}).  {\it For isovector excitations, where the neutrons move out of phase with the protons, one has an asymmetry of opposite sign to the elastic $A_{pv}\approx -A_{pv}^0$.}  I would expect excitations of mixed isospin to be in-between.  These estimates should also hold for giant resonances where for example the isovector giant dipole resonance should have $A_{pv}\approx -A_{pv}^0$.  I have calculated $A_{pv}$ for quasielastic scattering in ref. \cite{quasielastic} as discussed below.  

The measured asymmetry $A_{meas}$ includes contributions from both elastic scattering with asymmetry $A_{el}$ and from inelastic excitations,
\begin{equation}
A_{meas}=(1-f) A_{el}+ f\langle A_{in}\rangle\, .
\label{dilution}
\end{equation}
Here $f$ is the fraction of accepted events that involve an inelastic excitation of $^{27}$Al and $\langle A_{in}\rangle$ is the average parity violating asymmetry for these inelastic excitations.  This must be calculated and involves an appropriate cross section and acceptance weighted sum over the various inelastic excitations.   I invert Eq. \ref{dilution} to extract $A_{el}$ from $A_{meas}$,
\begin{equation}
A_{el}=\frac{A_{meas}-f\langle A_{in}\rangle}{1-f}\, .
\end{equation}
Now it is a simple matter to determine how accurately $f$ and $\langle A_{in}\rangle$ must be determined so that $A_{el}$ can be extracted with total error comparable to the experimental error in $A_{meas}$.  I define $\Delta A_f$ as the error in the extracted $A_{el}$ from an error $\Delta f$ in the inelastic fraction.  
\begin{equation}
\Delta A_f=\frac{A_{meas}-\langle A_{in}\rangle}{(1-f)^2}\, \Delta f 
\end{equation}     
Likewise $\Delta A_{in}$ is the error in $A_{el}$ from an error in $\langle A_{in}\rangle$,
\begin{equation}
\Delta A_{in}=\frac{f}{1-f}\, \Delta \langle A_{in}\rangle\, .
\end{equation}
I emphasize that $\Delta A_f$ will be small as long as $\Delta f$ is small.  I.E. $f$ is reasonably well determined from measured, or well known theoretical, cross sections and knowledge of the detector acceptance.  In this case $\Delta A_f$ will be small despite $\langle A_{in}\rangle$ having considerable uncertainty, or even differing in sign from $A_{meas}$.  Likewise $\Delta A_{in}$ will be small provided $f$ is small.  Indeed, I expect $f$ to be small, at low momentum transfers, became of the large elastic cross section.  For example suppose $f\approx 4\%$.  (Please note this is an arbitrary number.  The real value should be determined by the Qweak collaboration from studying the detector acceptance.)  In this case one would only need to determine $\langle A_{in}\rangle$ to 50\% in order for the error $\Delta A_{in}$ to be less than 2\%.    

Likewise $\Delta A_f$ will be less than 2\% provided that $\Delta f$ is determined to 25\% (of $f$).  This will be true even in the unfavorable case where $\langle A_{in}\rangle \approx -A_{meas}$.  This would require most of the inelastic strength to be isovector and is not expected.  If calculations give $\langle A_{in}\rangle$ small or positive than one would only need $f$ to 50\% for $\Delta A_f$ to be less than 2\%.  

The inelastic asymmetry $\langle A_{in}\rangle$ will have errors in the theoretical calculations of the parity violating asymmetries for various excited states.  In addition it will have errors from uncertainties in the relative contributions of different excited states.  {\it However as long as $f$ is relatively small, the required accuracy on $\langle A_{in}\rangle$ is very modest, perhaps of order 50\%}.  Therefore relatively crude calculations of the asymmetry may suffice.  

As an example I discuss calculations of the parity violating asymmetry for quasielastic scattering \cite{quasielastic}.  One does not need a detailed description of quasielastic scattering in order to determine $A_{pv}$ to of order 50\%.  Instead, all that is needed is a very rough idea of the ratio of isoscalar to isovector strength.   
      
Finally there are impurities in the Qweak target.  An alloy is used that is about 90\% $^{27}$Al but also contains some Cu, Mg, and Zn and other trace elements.  In future work I will calculate $A_{pv}$ for elastic scattering from these nuclei using relativistic mean field densities.

\section{Conclusions}
\label{Conclusions}

If the remaining nuclear structure uncertainty is in fact small and inelastic excitations can be estimated, {\it one can use the Qweak $A_{pv}$ data to confirm the neutron radius $R_n$ of $^{27}$Al}.  This is one of my main results.  To determine the sensitivity to $R_n$, I uniformly stretch the FSUgold neutron density, so that $R_n$ increases by 1\%, while keeping the proton density unchanged, see also \cite{Ban:2010wx}.  I then calculate the log derivative of $A_{DW}(C_0)$ with respect to $R_n$, evaluated at 7.8 degrees.  I find,
\begin{equation}
\Bigl|\frac{d \ln A_{DW}(C_0)}{d \ln R_n}\Bigr| \approx 2.
\end{equation}
This shows that a 4\% measurement of $A_{pv}$ is sensitive to 2\% changes in $R_n$.  {\it Therefore, in principle, one can use $A_{pv}$, good to 4\%, to probe the neutron radius of } $^{27}${\it Al  to about 2\%.} For comparison, the PREX experiment measured $R_n$ of $^{208}$Pb to 3\% \cite{ref:prexI}.  Note that the follow up experiment PREX-II aims to improve this to 1\% \cite{ref:prexII}.

If one estimates contributions from inelastic excitations and impurity scattering to the necessary accuracy, and the nuclear structure uncertainties for $^{27}$Al are indeed small, one can determine $R_n$.  What is the physics content of this measurement?  For $^{208}$Pb, $R_n$ determines the density dependence of the symmetry energy and the pressure of pure neutron matter with important applications to astrophysics \cite{bigprex,cjhjp_prl,roca-maza,rNSvsRn}.  However for $^{27}$Al, $N\approx Z$ and many relativistic mean field models have $R_n\approx R_p$, where $R_p$ is the proton radius.  There may only be a small range of $R_n$ values predicted by all reasonable nuclear structure models.  This should be explicitly checked by looking at a large number of nuclear structure models.  Very likely theory makes a sharp prediction for $R_n-R_p$ for $^{27}$Al that is essentially independent of the density dependence of the symmetry energy.  Therefore the measurement of $R_n$ could provide a sharp test of theory and experiment.  A disagreement would suggest an important problem either in the measurement or in the theory.  While agreement of the Qweak $R_n$ measurement with theory would support the validity of many aspects of the measurement, analysis, and theory.

%\section{Results}\label{Results}

%\section{Conclusions}\label{Conclusions}

\begin{acknowledgments}
Robert Michaels and Rupesh Silwal are thanked for helpful discussions and for information on the Qweak $^{27}$Al measurement.   This research was supported in part by DOE grants DE-FG02-87ER40365 (Indiana University) and DE-SC0008808 (NUCLEI SciDAC Collaboration).
\end{acknowledgments}

\end{document}